\documentclass[twocolumn, times, twocolappendix]{aastex63}
\newcommand{\cione}{[\ion{C}{1}]\,}

\received{}
\revised{}
\accepted{}
\submitjournal{}

\shorttitle{Atomic carbon in a z=2.49 protocluster}
\shortauthors{Minju M. Lee et al.}
\defcitealias{minju2017a}{Paper I}
\defcitealias{minju2019a}{Paper II}
\defcitealias{Valentino2020b}{V20}

\begin{document}

\title{Revisited cold gas content with atomic carbon [\ion{C}{1}] in z=2.5 protocluster galaxies}

\correspondingauthor{Minju M. Lee}
\email{minju@mpe.mpg.de}

\author[0000-0002-2419-3068]{Minju M. Lee}
\affiliation{Max-Planck-Institut f\"{u}r Extraterrestrische Physik (MPE), Giessenbachstr., D-85748 Garching, Germnay}

\author[0000-0002-4937-4738]{Ichi Tanaka}
\affiliation{Subaru Telescope, National Astronomical Observatory of Japan, 650 North Aohoku Place, Hilo, HI 96720, USA}

\author[0000-0002-2364-0823]{Daisuke Iono}
\affiliation{National Astronomical Observatory of Japan, 2-21-1 Osawa, Mitaka, Tokyo 181-8588, Japan}
\affiliation{The Graduate University for Advanced Studies (SOKENDAI),2-21-1 Osawa, Mitaka, Tokyo 181-8588, Japan}

\author[0000-0002-8049-7525]{Ryohei Kawabe}
\affiliation{National Astronomical Observatory of Japan, 2-21-1 Osawa, Mitaka, Tokyo 181-8588, Japan}
\affiliation{The Graduate University for Advanced Studies (SOKENDAI),2-21-1 Osawa, Mitaka, Tokyo 181-8588, Japan}

\author[0000-0002-2993-1576]{Tadayuki Kodama}
\affiliation{Astronomical Institute, Tohoku University, Aoba-ku, Sendai 980-8578, Japan}

\author[0000-0002-4052-2394]{Kotaro Kohno}
\affiliation{Institute of Astronomy, Graduate School of Science, University of Tokyo, 2-21-1 Osawa, Mitaka, Tokyo 181-0015, Japan}
\affiliation{Research Center for the Early Universe, School of Science, The University of Tokyo, 7-3-1 Hongo, Bunkyo-ku, Tokyo 113-0033, Japan}

\author[0000-0002-2501-9328]{Toshiki Saito}
\affiliation{Max-Planck Institute for Astronomy, Konigstuhl 17, D-69117, Heidelberg, Germany}

\author[0000-0003-4807-8117]{Yoichi Tamura}
\affiliation{Department of Physics, Nagoya University, Furo-cho, Chikusa-ku, Nagoya 464-8601, Japan}




\begin{abstract}
We revisit the cold gas contents of galaxies in a protocluster at $z=2.49$ using the lowest neutral atomic carbon transition \cione$\,^3P_1-^3P_0$ from Atacama Large Millimeter/submillimeter Array observations. We aim to test if the same gas mass calibration used in field galaxies can be applied to protocluster galaxies.
Five galaxies out of sixteen targeted galaxies are detected in the \cione line, and these are all previously detected in CO~(3--2) and CO~(4--3) and three in 1.1 mm dust continuum.
We investigate the line luminosity relations between CO and \cione in the protocluster and compare with other previous studies.
We then compare the gas mass based on three gas tracers of \cione, CO(3--2), and dust if at least one of the last two tracers are available. 
Using the calibration adopted for field main-sequence galaxies, the \cione-based gas measurements are lower than or comparable to the CO-based gas measurements by -0.35 dex at the lowest with the mean deviation of -0.14 dex.
The differences between \cione- and the dust- based measurements are relatively mild by up to 0.16 dex with the mean difference of 0.02 dex.
Taking these all together with calibration uncertainties, with the \cione~line,  we reconfirm our previous findings that the mean gas fraction is comparable to field galaxies for a stellar-mass range of $\log{(M_{\rm star}/M_{\odot})} = [10.6, 11.3]$. 
However, at least for these secure five detections, the depletion time scale decreases more rapidly with stellar mass than field galaxies that might be related to earlier quenching in dense environments.

\end{abstract}

\keywords{galaxies: clusters: general -- galaxies: evolution -- galaxies: high-redshift -- galaxies: ISM -- large-scale structure of universe}


\section{Introduction} \label{sec:intro}
Molecular gas plays a prominent role in the evolution of galaxies by fuelling star-forming activities.
The bulk of the gas is the molecular hydrogen (H$_2$), but it does not have a permanent dipole moment.  
A weak emission is only allowed at a high energy level, hampering direct observations of it in the typical ISM condition.
As alternative tracers, CO lines and dust have been used to weigh the total molecular gas in galaxies through the ``conversion factors" calibrated by the Milky Way and nearby galaxy observations.
With the advent of sensitive interferometric facilities such as the Atacama Large Millimeter/submillimeter Array (ALMA), the NOrthern Extended Millimeter Array (NOEMA), and the Karl G. Jansky Very Large Array (JVLA), the cosmic evolution of molecular gas up to $z\sim6$ is measured for many field galaxies (e.g., \citealt{Tacconi2020}). 

However, it is still an open question whether the same calibration (or conversion factor) can be applied to galaxies in dense environments of clusters or their progenitors, protoclusters, because the calibration might be different from isolated field galaxies due to different excitation conditions, metallicity, and ambient radiation fields.
Such a test is necessary by using an independent method, whether the calibration of field galaxies can be applied in the same manner. 

Fine structure lines of atomic carbon [\ion{C}{1}] is an alternative tracer of the molecular gas.
A traditional theoretical model of photodissociation regions (PDRs) viewed that the \cione line is originated from a narrow region at the interface between [\ion{C}{2}] and CO (e.g., \citealt{Tielens1985a, Tielens1985b, Hollenbach1999, Kaufman1999}), 
but it was challenged by the observations that CO and \cione can coexist based on the constant column density ratio N(\cione)/N(CO) over a wide range of ambient FUV field and physical conditions in the Orion clouds and Galactic center (\citealt{Ikeda1999,Ikeda2002,Ojha2001}) 
pointing to a potential use of \cione as a molecular gas tracer (\citealt{Gerin2000, Papadopoulos2004a}).
A tight, linear correlation between the CO~(1--0) and \cione line luminosities over five orders of magnitude for local galaxies strengthened this view (e.g., \citealt{Jiao2019}).

Not only is an independent molecular gas tracer, but \cione line can also be an efficient gas tracer than the lowest CO~(1--0) in terms of observing time with the current facilities. 
The current JVLA can target CO (1--0) line for $z>1.3$ galaxies, observing the \cione can be more efficient than CO~(1--0) considering the positive K-correction at the same resolution and the robustness of \cione\,against CMB at very high redshifts ($z\gtrsim4$, \citealt{daCunha2013, Zhang2016}).
The frequency coverage of ALMA and NOEMA only allows $J>1$ transitions of CO for galaxies at $z\gtrsim0.5$ so that one needs an additional assumption of gas excitation to convert the line flux into the CO~(1--0) luminosity.

The lowest transition of \cione $^3P_1-$$^3P_0$ (hereafter written simply as the \cione (1--0) transition) is optically thin in most cases, less sensitive to excitation temperature compared to \cione $^3P_2-$$^3P_1$ (hereafter \cione (2--1)) when converting to a mass (\citealt{Weiss2005}).
Similar to CO~(1--0)-to-H$_2$ conversion factor, converting the \cione (1--0) line flux into a gas measurement needs an assumption of the \cione/[H$_2$] abundance ratio (hereafter $X_{\rm CI}$).
Recent calculations showed that $X_{\rm CI}$ is robust or well-behaved even in low metallicity and high cosmic ray environments (e.g., \citealt{Offner2013, Glover2016, Bisbas2017}).

So far, \cione line searches and detections at high-$z$ have happened in galaxies well above the star-forming main-sequence, or on the main-sequence in general fields (e.g., \citealt{Walter2011, Alaghband-Zadeh2013, Popping2017, Valentino2018, Valentino2020b} (hereafter V20); \citealt{Bourne2019}).
In this Paper, we perform a [CI] line search by targeting main-sequence galaxies associated with a protocluster at $z=2.49$, where CO and dust emissions are already detected (\citealt{minju2017a, minju2019a}, hereafter we refer to these two papers as \citetalias{minju2017a} and \citetalias{minju2019a}, respectively).
These main-sequence star-forming galaxies are H$\alpha$ emitters selected by a narrow-band filter technique using the Subaru telescope (\citealt{Tanaka2011}).
We report \cione(1--0) detections using ALMA and revisit the gas mass and compare gas masses derived from the three different tracers.

This paper is organized as follows.
In Section~\ref{sec:obs}, we present a summary of the ALMA Band 4 observations targeting the \cione line.
In Section~\ref{sec:results}, we show the \cione line detections and compare the line properties with the previous CO observations.
We discuss the derived molecular mass from different gas tracers in Section~\ref{sec:discussion}.
Finally, a summary is presented in Section~\ref{sec:summary}.
Throughout this paper, we assume a $\Lambda$ cold dark matter cosmology with $H_0$ =70 km s$^{-1}$ Mpc$^{-1}$, $\Omega _{\rm m}$ = 0.3, and $\Omega_{\Lambda}=0.7$. 
The adopted initial mass function (IMF) is the Chabrier initial mass function (IMF; \citealt{Chabrier2003}) in the mass range of 0.1-100 $M_{\odot}$.

\section{Observations} \label{sec:obs}
\subsection{ALMA Band 4 observations} \label{sec:obsdetail}
We used ALMA Band~4 receivers to obtain the redshifted \cione~(1--0) emission (ID: ADS/JAO.ALMA \#2015.1.00152.S, PI: Minju Lee).
We designed our observations to use two-point positions, which is the same as the CO~(4--3) observations summarized in \citetalias{minju2019a}, covering a total of 16 out of 25 parent H$\alpha$ emitters (HAEs).
A total of either 38 or 39 antennas were used with the baseline length between 15 m and 3.1 km. 
For each pointing, on-source time was $\simeq$95 mins to get the line emission.
We used four spectral windows (SPWs), two of each were placed in the upper and the lower sideband, respectively.
One SPW was set in the Frequency-Division Mode (FDM) to detect the redshifted \cione (1--0) with a channel width of 7.82 MHz ($\sim$ 16.6 km s$^{-1}$) covering 1.875 GHz bandwidth.
The remaining three SPWs were observed in the Time-Division Mode (TDM) with a 2.0-GHz bandwidth at the 15.6-MHz resolution to cover the dust continuum at 2 mm.
J2025+3343 and J2148+0657 were chosen for bandpass calibrators.
J2148+0657 and J1751+0939 were used for flux calibration. 
A phase calibrator was J2114+2832.

We used $\mathtt{CASA}$ (\citealt{McMullin2007}) version 4.7.0 and 5.6.1 for the calibration of visibility data and imaging, respectively.
For visibility calibration, we used the pipeline script provided by the ALMA Regional Center staffs.

Images were produced by $\mathtt{CASA}$ task, $\mathtt{tclean}$, and deconvolved down to 2$\sigma$ noise level.
The CLEAN masks were created based on the positions of HAEs taking $1''$-radius circular regions.
The synthesized beam is $0^{\prime\prime}.42\times0^{\prime\prime}.29$ and the typical noise level is 0.10 mJy beam$^{-1}$ around the phase center at 80 km s$^{-1}$ for natural weighting.
We take the velocity resolution of 80 km s$^{-1}$ as a default, or otherwise specified.
Tapered images were also created by three different uvtaper parameters of $0''.3$, $0''.5$ and $0''.9$  in the $\mathtt{tclean}$ task to check the robustness of the line detections and the presence of extended emissions.
The corresponding beam sizes are $0''.53 \times 0''.43$, $0''.65 \times 0''.57$ and $0''.97 \times 0''.92$ for the uvtaper values of $0''.3$, $0''.5$ and $0''.9$, respectively.
We subtracted the continuum emission on the image domain using $\mathtt{imcontsub}$ (fitorder = 0) after we made smaller data cubes ($10''\times10''$) centered on the target HAEs.
This is to get improved results of continuum subtraction for galaxies not close to the phase center.
We then performed a primary beam correction.
Hereafter, we refer to the data cube that is corrected for the primary beam response and of which the continuum emission is subtracted.

\subsection{Detection criteria and flux measurements}\label{sec:criteria}
We impose the same detection criteria applied in the CO~(4--3) line detection (\citetalias{minju2019a}) that were found to be robust to identify line detections for known positions.
In summary, we regard a galaxy as detected by imposing double-step criteria. 
Firstly, the \cione line detection candidates are selected if (a) a signal-to-noise (S/N) ratio at the peak position ($I_{\rm CI,peak}$) is equal to or greater than 4.5 ($S/N_{I_{\rm CI,peak}}\geq 4.5$). 
We then regard these candidates as detected if the peak position spectrum satisfies either of the following criteria : (b) a peak flux density ($S_{\rm CI, peak}) \geq 3.5~\sigma$;
(c) at least two continuous channels, including a maximum-peak-flux channel have fluxes $> 2.5~\sigma$, where $\sigma$ is the average channel noise level, estimated from the line-free positions.

The noise estimate for the peak line intensity is based on the calculation after taking into account the channel noise, integrating channel range following \cite{Hainline2004} assuming the source is unresolved.
We also checked the noise level by bootstrapping from 30 random positions of the map, which matches the theoretical expectation.
The total line flux is measured by performing 2D gaussian fitting using $\mathtt{imfit}$ in CASA for a given aperture.

We check the growth curves in natural and tapered maps to find the optimal imaging parameters that give the highest S/Ns and total flux.
In summary, for HAE~3 and HAE~16, we use the $0''.3$-tapered map and a circular aperture size of $1''.0$.
For HAE~4, we use the natural-weighted map and a circular aperture size of $0''.7$ (but see Section~\ref{sec:det} for a different choice of aperture size for ``HAE~4b''). 
For HAE~8 and HAE~9, we use the $0''.5$-tapered map and an aperture size of $1''.0$. 

For HAE~8, we use 50 km s$^{-1}$ considering the narrow width known from the previous CO line detections.
For low S/N galaxies (HAE~4 and HAE~3), we also checked whether different velocity binnings (50 km s$^{-1}$ versus 80 km s$^{-1}$) change our results. We confirmed that the results did not change in terms of the line widths and fluxes within uncertainties between two different velocity binnings.

For \cione(1--0) non-detection, we calculate a 3$\sigma$ limit for galaxies by assuming the same line width and aperture sizes used in the CO~(4--3) if it is detected, and in other cases, we take a full-width half-maximum (FWHM) of 300 km $^{-1}$ for the integrated flux.
We note that except for HAE~5, which was not targeted for the \cione(1--0) and CO~(4--3) observations, galaxies with CO~(3--2) detection are a subset of the CO~(4--3) detected galaxy group, i.e., if CO~(3--2) line is detected, then CO~(4--3) is also detected. If a galaxy is not detected in any of the lines, we use the natural weight image and the aperture size of $0''.7$ for calculating the \cione flux limit.
The current detection criteria exclude a possible \cione detection with the {\it peak} S/N = 5.4 for HAE~7, which was previously detected in CO~(4--3), but not satisfying neither (b) nor (c) in Step 2 for the \cione analysis (the spectra are presented in Appendix~\ref{app:single}).
The 2D Gaussian best-fit flux ($0.28\pm0.09$) gives $\approx$10\% higher flux than the 3$\sigma$ upper limit placed by the noise level, but still within the uncertainty and it does not impact our later discussion.
The data cubes were also visually checked further for the candidate galaxies without any CO line detections because of the redshift uncertainty, which is $\Delta z\approx0.03$, coming from the original selection (i.e., narrow-band filter detected).

\subsection{Reanalysis of previous CO and dust measurements}\label{sec:reanalysis}
We reanalyzed the CO~(3--2), CO~(4--3), and dust data for unified analyses to reduce the systematic errors coming from different flux measurement methods.
The flux values in our previous studies were measured in different spatial and/or spectral resolutions and various methods.
Here we performed the same analyzing procedures applied in the measurement of the \cione line.
The beamsizes of the CO~(4--3), CO~(3--2) and dust are $0''.53 \times 0''.32$, $0''.89\times0''.66$, and $0''.78\times 0''.68$, respectively, for natural weighting. 
As the beam size of the CO~(4--3) observations is very similar to that of \cione observations, we took the same aperture sizes and imaging parameters described in Section~\ref{sec:criteria} when doing $\mathtt{imfit}$.
We used the natural weighting map for CO~(3--2) and for dust continuum, taper parameter of $0''.5$, which gives the synthesized beam of $0''.89\times0''.82$.
For the CO~(3--2) data and dust, we used a fixed aperture size of $1''.0$ to measure the flux values.

The choice of the aperture sizes in flux measurement is based on our investigation to maximize the S/N. 
However, we note that choosing $1''.0$ for all galaxies instead of different aperture sizes does not change our flux estimate.
Indeed, HAE~4 is the only galaxy using different aperture sizes between different lines, we note that \cione and CO~(4--3) fluxes using the $1''.0$-circular aperture for HAE~4 are consistent with the $0''.7$ measurement but with low S/N (but see Section~\ref{sec:det} for further discussion on HAE~4). 
Since we probe close transitions between $J=3$ and $4$ for CO, the difference in radial distribution might be small. 
We note, however, that for non-detection different aperture sizes can give $\approx40\%$ (or $\approx 0.14$ dex) change in the 3$\sigma$ upper limit.

The number of line detection does not change (i.e., seven galaxies for CO~(3--2), eleven galaxies for CO~(4--3), and four galaxies for dust), but the total flux values and signal-to-noise ratios are updated.
For CO~(4--3), HAE~3 and HAE~16 fluxes increased by $\approx40-50\%$ than \citetalias{minju2019a}. We also note that the S/Ns of HAE~1, 2, 7, 12 and 23, which are detected only in CO~(4--3), are lower than the previous measurements that we decided not to include in the line ratio analysis in Section~\ref{sec:lumiratio} (see their spectra in Appendix~\ref{app:single}).
The CO~(3--2) fluxes of HAE~8 and HAE~4 have increased roughly by a factor of two than the one in \citetalias{minju2017a}, where we took a peak flux from the smoothed maps, instead of fitting Gaussian for all galaxies (see the Appendix for the related discussions on the flux measurements in \citetalias{minju2017a}).
However, we note that the updated values are consistent with the values reported in \citet{Tadaki2019a} within uncertainties.

The updated fluxes do not change the main conclusion made in \citetalias{minju2017a} that (1) the mean gas fraction of galaxies with $\log{(M_{\rm star}/M_{\odot})} = [10.6, 11.3]$ is consistent with the field within the error bars and (2) the depletion time scale (or star-formation efficiency) decreases (increases) with increasing stellar mass but with an outlier of an AGN-dominated galaxy (HAE 5).
We summarize the flux values in Table~\ref{tab:lineflux}. Table~\ref{tab:lineflux} supersedes the previous measurements and we will regard the updated measurements for further analysis.

\section{Results} \label{sec:results}
\subsection{Detection of \cione and line properties compared to CO}\label{sec:det}
\begin{figure*}[tbh]
\centering
\includegraphics[width=0.85\textwidth, bb=0 0 1400 1800]{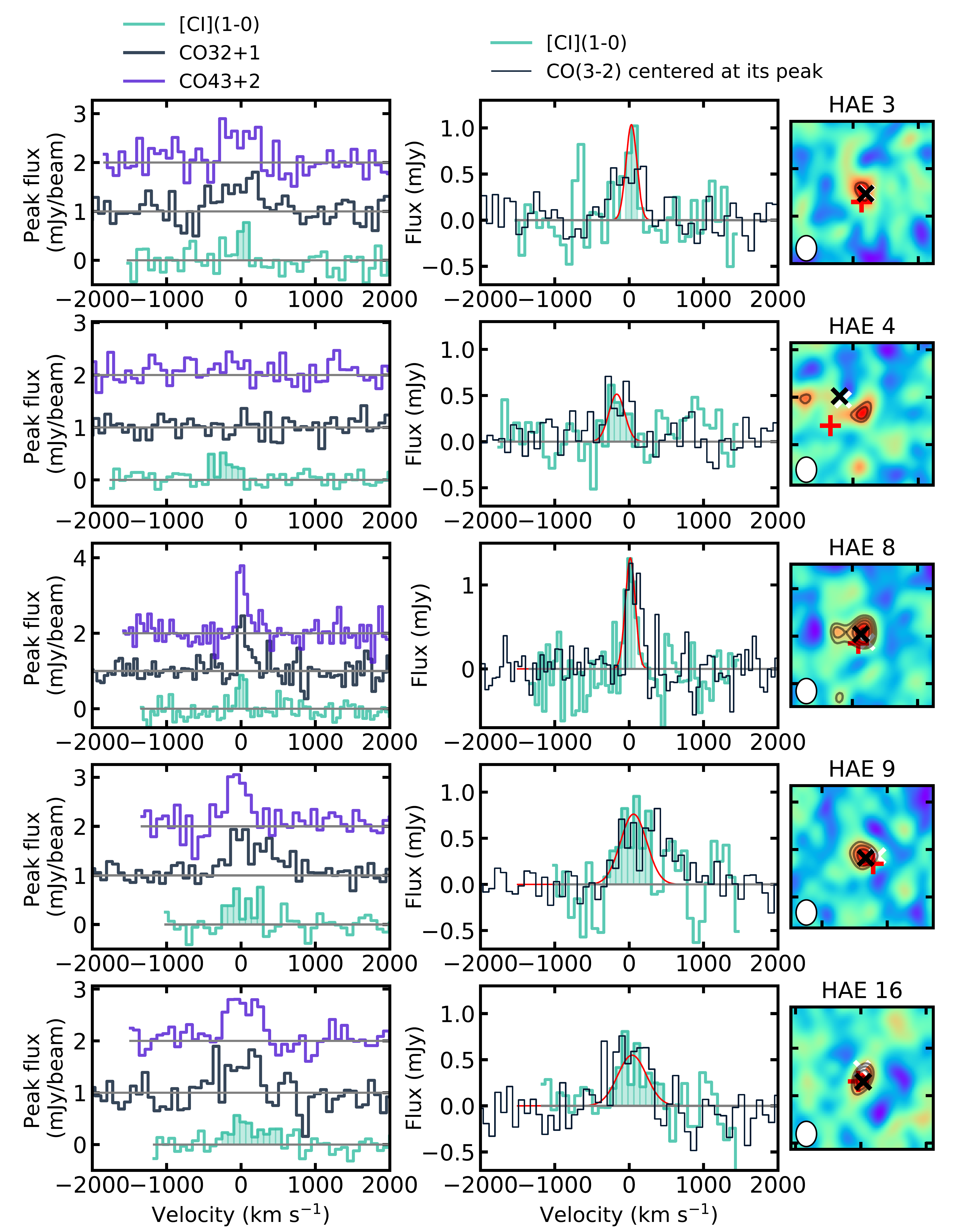}
\caption{Left: \cione (1--0) line spectrum for detections at the \cione peak. Color-filled regions indicate the velocity range to integrate to derive the line flux for \cione (1--0), which gives the largest S/N. For comparison, CO~(4--3) and CO~(3--2) spectrum at the \cione peak position are shown with a base level shift of 2.0 and 1.0, respectively. The velocity resolution is 80 km s$^{-1}$ except for HAE~8. For HAE~8, the resolution is 50 km s$^{-1}$. Middle: \cione (1--0) line spectrum using a $1''.0$ aperture centered at the peak. CO~(3--2) spectrum using the same aperture size at the CO~(3--2) peak is plotted in black thinner line. Right: the velocity-integrated map of \cione (1--0). The size of the panel is $3''\times 3''$. Cross symbols indicate the peak positions of CO~(4--3) (black), CO~(3--2) (red), and H$\alpha$ (white) emissions. The contour levels are starting from $4\sigma$ in steps of $1\sigma$. Dashed contours are also plotted for $-4\sigma$ indicating no 4-$\sigma$ peak around the galaxies. The synthesized beam ($0''.53\times 0''.43$) is shown as an filled-ellipse on the bottom left of each panel. \label{fig:lines}}
\end{figure*}

We detect the \cione line from five galaxies previously detected in CO~(3--2) and CO~(4--3) lines, i.e., HAE~3, 4, 8, 9 and 16.
Three galaxies, HAE~3, 8, and 9, were detected in the 1.1 mm continuum data as well.
In the left panels of Figure~\ref{fig:lines}, we show the spectra of \cione, CO~(3--2), and CO~(4--3) emissions at the peak position of the \cione(1--0) emission for [CI] detections.
The spectra of the \cione and CO~(4--3) lines are taken from the $0''.3$-tapered maps and CO~(3--2) from the natural-weighted maps.

We find the line properties (widths and redshifts) of \cione centered at each peak position of the \cione and CO~(3--2) with a fixed aperture ($1''.0$) size are broadly consistent with those of CO~(3--2), except for HAE~3 and HAE~9 (Figure~\ref{fig:lines} middle panels).
These two galaxies show signatures of mergers in their stellar and/or CO maps (\citetalias{minju2017a,minju2019a}).
For HAE~3, the \cione line is narrower (FWHM = 172 $\pm$ 106 km s$^{-1}$) than two CO lines (FWHM =  500 km s$^{-1}$) which may indicate kinematically different origins of the emission lines. If we integrate the [C I] cube of HAE~3 over the same velocity range defined by two CO lines, the 1D Gaussian fitted flux ($0.15\pm0.08$) is consistent within uncertainties.
For HAE~9, we discussed its kinematic properties in \citetalias{minju2019a} that the galaxy is undergoing a gas-rich merger.
The red-shifted component of HAE~9 is not detected in \cione (or does not satisfy the detection criteria), which was detected in CO~(4--3).
Accordingly, in Table~\ref{tab:lineflux}, we list additional flux measurements for HAE~3 and HAE~9 (named HAE~3a and HAE~9a) to refer to the flux values for which integrating ranges are matched to those of the CO~(3--2) lines. In the later analysis of the gas measurements, we will use these values.

The peak positions of \cione are also consistent within the positional errors except for HAE~4 if the beam size and the S/Ns are taken into account (Figure~\ref{fig:lines} right panels).
HAE~4 shows a positional offset of $0.6''$ between H$\alpha$ and CO~(3--2) peak positions (\citetalias{minju2017a}) and extended H$\alpha$ emission.
The CO~(4--3) peak is closer to the H$\alpha$ peak in contrast to the offset from CO~(3--2) peak position.
The \cione line peak is also offset from the other line peak positions by $\approx 0''.5-0''.6$ ($\approx$ 4-5 kpc).
It may indicate different gas excitation conditions within the galaxy as discussed in \citetalias{minju2019a}.
For this reason, we take another flux measurement for HAE~4. 
We list the line fluxes from different positions which give the highest S/N for individual emission lines (HAE~4), and from the CO~(4--3) peak position (HAE~4b, see Appendix~\ref{app:haes} for its spectrum). 
For HAE~4b, we derive the fluxes using a $1''.0$ circular aperture to be consistent with other \cione detected galaxies. 
Owing to the low S/N at the CO~(4--3) position, we do not derive the flux in the 2D map for HAE~4b using $\texttt{imfit}$ task in CASA, but by fitting the 1D spectrum with a single Gaussian. 
We note that the aperture photometry in the 2D map, using the same circular aperture of $1''.0$, is lower than or comparable to the 1D Gaussian fit listed in Table~\ref{tab:lineflux}, i.e., $0.13\pm0.03$, $0.14\pm0.04$, $0.20\pm0.05$ for \cione, CO~(3--2) and CO~(4--3), respectively.
Other galaxies also show similar signatures in their CO and \cione line profiles; the line profiles taken from the the \cione peak position show a varying degree of relative line ratios at each velocity bin (Figure~\ref{fig:lines} left panels).
Discussing the resolved gas excitation conditions is beyond the scope of this paper, and this would need more sensitive and higher angular resolution observations to get more details.
In this paper, we assume that the line emissions are {\it globally} coming from the same galaxy.

\begin{deluxetable*}{ccccccccccc}
\tablecaption{{Summary of line observations for HAEs associated to 4C23.56 protocluster}\label{tab:lineflux}}
\tabletypesize{\footnotesize}
\tablewidth{0.99\textwidth}
\tablehead{
\colhead{ID} & \colhead{R.A.}  & \colhead{Decl.} 	&	\colhead{\cione beam}	&\colhead{aperture([CI])}	& \colhead{$\Delta v({\rm [C I]})_{\rm integ.}$}	&	\colhead{FWHM$_{\rm [CI](1-0)}$} & \colhead{$I_{\rm [CI]10}$} &\colhead{$I_{\rm CO32}$} 	&  \colhead{$I_{\rm CO43}$}   & \colhead{$S_{\rm 1.1mm}$}	\\
\colhead{} & \colhead{J2000}  & \colhead{J2000} & \colhead{$''\times''$} & \colhead{$''$} & \colhead{km s$^{-1}$} & \colhead{km s$^{-1}$}	& \colhead{Jy km s$^{-1}$}& \colhead{Jy km s$^{-1}$} & \colhead{Jy km s$^{-1}$} &  \colhead{mJy} \\
\colhead{(1)} & \colhead{(2)}  & \colhead{(3)} &\colhead{(4)} & \colhead{(5)} & \colhead{(6)} &  \colhead{(7)} & \colhead{(8)} & \colhead{(9)}	& \colhead{(10)} & \colhead{(11)}}
\startdata
HAE1\tablenotemark{*}		&	316.81166	&	23.52921	&	0.42$\times$0.29	&	0.7	&	560	&	...	&	$<0.17$	&	$<0.18$	&	0.34$\pm$0.13	&	$<0.25$\\
HAE2\tablenotemark{*}		&	316.84074	&	23.53043	&	0.42$\times$0.29	&	0.7	&	400	&	...	&	$<0.14$	&	$<0.12$	&	0.16$\pm$0.07	&	$<0.36$\\
HAE3		&	316.83763	&	23.52055	&	0.53$\times$0.43	&	1.0	&	160	&	172$\pm$106	&	0.17$\pm$0.05	&	0.43$\pm$0.12	&	0.67$\pm$0.17	&	0.77$\pm$0.26\\
HAE3a\tablenotemark{$\dagger$}		&	316.83763	&	23.52055	&	0.53$\times$0.43	&	1.0	&	560	&	172$\pm$106	&	0.15$\pm$0.08	&	0.43$\pm$0.12	&	0.63$\pm$0.18	&	0.77$\pm$0.26\\
HAE4		&	316.84013	&	23.52810	&	0.42$\times$0.29	&	0.7	&	480	&	228$\pm$130	&	0.20$\pm$0.06	&	0.42$\pm$0.10	&	0.19$\pm$0.06	&	$<0.42$\\
HAE4b\tablenotemark{$\ddag$}		&	316.84013	&	23.52810	&	0.53$\times$0.43	&	1.0	&	480	&	292$\pm$138	&	0.13$\pm$0.08	&	0.17$\pm$0.07	&	0.22$\pm$0.10	&	$<0.42$\\
HAE5		&	316.82069	&	23.50846	&	...	&	...	&	...	&	...	&	...	&	0.14$\pm$0.04	&	...	&	$<0.29$\\
HAE6		&	316.83955	&	23.52209	&	0.42$\times$0.29	&	0.7	&	300	&	...	&	$<0.09$	&	$<0.14$	&	$<0.11$	&	$<0.31$\\
HAE7\tablenotemark{*}		&	316.81468	&	23.52707	&	0.53$\times$0.43	&	1.0	&	1520	&	...	&	$<0.25$	&	$<0.23$	&	0.43$\pm$0.12	&	$<0.51$\\
HAE8		&	316.81638	&	23.52431	&	0.65$\times$0.57	&	1.0	&	150	&	154$\pm$62	&	0.24$\pm$0.04	&	0.55$\pm$0.09	&	0.49$\pm$0.06	&	0.90$\pm$0.19\\
HAE9		&	316.84409	&	23.52871	&	0.65$\times$0.57	&	1.0	&	560	&	410$\pm$171	&	0.38$\pm$0.09	&	0.64$\pm$0.10	&	0.61$\pm$0.12	&	1.52$\pm$0.29\\
HAE9a\tablenotemark{$\dagger$}		&	316.84409	&	23.52871	&	0.65$\times$0.57	&	1.0	&	800	&	410$\pm$171	&	0.55$\pm$0.14	&	0.64$\pm$0.10	&	1.23$\pm$0.23	&	1.52$\pm$0.29\\
HAE10		&	316.81543	&	23.52003	&	0.53$\times$0.43	&	1.0	&	480	&	...	&	$<0.17$	&	0.53$\pm$0.12	&	0.26$\pm$0.10	&	0.50$\pm$0.16\\
HAE11		&	316.79338	&	23.51373	&	...	&	...	&	...	&	...	&	...	&	.....	&	...\\
HAE12\tablenotemark{*}		&	316.81222	&	23.52988	&	0.42$\times$0.29	&	0.7	&	240	&	...	&	$<0.13$	&	$<0.12$	&	0.19$\pm$0.07	&	$<0.25$\\
HAE13		&	316.84092	&	23.52826	&	0.42$\times$0.29	&	0.7	&	300	&	...	&	$<0.09$	&	$<0.10$	&	$<0.11$	&	$<0.32$\\
HAE14		&	316.83241	&	23.51417	&	...	&	0.7	&	300	&	...	&	...	&	$<0.11$	&	...	&	$<1.05$\\
HAE15		&	316.83315	&	23.51896	&	...	&	0.7	&	300	&	...	&	...	&	$<0.10$	&	...	&	...\\
HAE16		&	316.81102	&	23.52107	&	0.53$\times$0.43	&	1.0	&	800	&	453$\pm$201	&	0.33$\pm$0.07	&	0.57$\pm$0.14	&	0.82$\pm$0.12	&	$<0.94$\\
HAE17		&	316.82340	&	23.53068	&	...	&	...	&	300	&	...	&	...	&	$<0.20$	&	...	&	...\\
HAE18		&	316.84011	&	23.53366	&	...	&	...	&	300	&	...	&	...	&	$<0.14$	&	...	&	...\\
HAE19		&	316.84247	&	23.52944	&	0.42$\times$0.29	&	0.7	&	300	&	...	&	$<0.11$	&	$<0.10$	&	$<0.13$	&	$<0.27$\\
HAE20		&	316.81228	&	23.52238	&	0.42$\times$0.29	&	0.7	&	300	&	...	&	$<0.08$	&	$<0.13$	&	$<0.10$	&	...\\
HAE21		&	316.81175	&	23.52857	&	0.42$\times$0.29	&	0.7	&	300	&	...	&	$<0.11$	&	$<0.13$	&	$<0.13$	&	$<0.26$\\
HAE22		&	316.82441	&	23.52909	&	...	&	...	&	300	&	...	&	...	&	$<0.20$	&	...	&	...\\
HAE23\tablenotemark{*}		&	316.81147	&	23.52184	&	0.42$\times$0.29	&	0.7	&	800	&	...	&	$<0.13$	&	$<0.23$	&	0.27$\pm$0.09	&	...\\
HAE24		&	316.84344	&	23.54678	&	...	&	...	&	...	&	...	&	...	&	.....	&	...\\
HAE25		&	316.80971	&	23.53572	&	...	&	...	&	...	&	...	&	...	&	.....	&	...\\
\enddata
\tablecomments{\\
$\dagger$: flux values obtained by integrating the velocity range used in CO~(3--2) line (column 6) \\$\ddagger$: flux values at a fixed position to the CO~(4--3) peak, which is the closest ($\Delta r= 0''.15$) to the original H$\alpha$ emission. \\
$*$: HAE~1, 2, 7, 12, and 23 satisfy our detection criteria in CO~(4--3) but we note that they have lower significance levels of the line detection in comparison with other secure detections those have other line and/or dust detections (see Appendix~\ref{app:single} for their spectra). Therefore, we suggest readers use these values with caution.\\ 
Colums: (1) Source ID. For sources without a dagger mark($\dagger$) or a double dagger mark ($\ddag$) behind the ID, we list fluxes integrated over the velocity range with beams and aperture sizes that maximizes the S/N of the peak position flux and total flux, centered at the peak position of each line; (2) Right Ascension (J2000) of H$\alpha$ position from Subaru/MOIRCS observations (\citealt{Tanaka2011}; I. Tanaka in preparation); (3) Declination (J2000) of H$\alpha$ position; (4) Synthesized beam of the \cione images to measure the flux; (5) Aperture size to measure the \cione flux or the 3$\sigma$ upper limit;  (6) Velocity integrating range to measure the \cione flux. If the \cione line is non-detection but other lines (i.e., CO~(4--3) or CO~(3--2)) are detected, we used the integrating range referring to their integrating ranges that is listed in this column. If no other lines are detected, we set the integrating range fixed as 300 km s$^{-1}$ for 3$\sigma$ upper limits; (7) Full-width half-maximum of the \cione line fitted with a single 1D Gaussian profile; (8) \cione (1--0) flux measured using CASA $\texttt{imfit}$ (but 1D Gaussian fit for HAE~4b) with the aperture size listed in Col. (5) and (6). Upper limits are 3$\sigma$; (9) CO~(3--2) flux using CASA $\texttt{imfit}$ (but 1D Gaussian fit for HAE~4b) with an aperture of $1''.0$. Upper limits are 3$\sigma$ using the value listed in Col (5) and (6); (10) CO~(4--3) flux using CASA $\texttt{imfit}$ (but 1D Gaussian fit for HAE~4b) with the same aperture size listed in Col. 5. Upper limits are 3$\sigma$ using the value listed in Col (5) and (6); (11) 1.1 mm dust continuum flux using an aperture of $1''.0$. All measurements presented here supersede the previous measurement reported in \citetalias{minju2017a} and \citetalias{minju2019a}.}
\end{deluxetable*}

\begin{figure}[tb]
\centering
\includegraphics[width=0.42\textwidth, bb=0 0 900 1800]{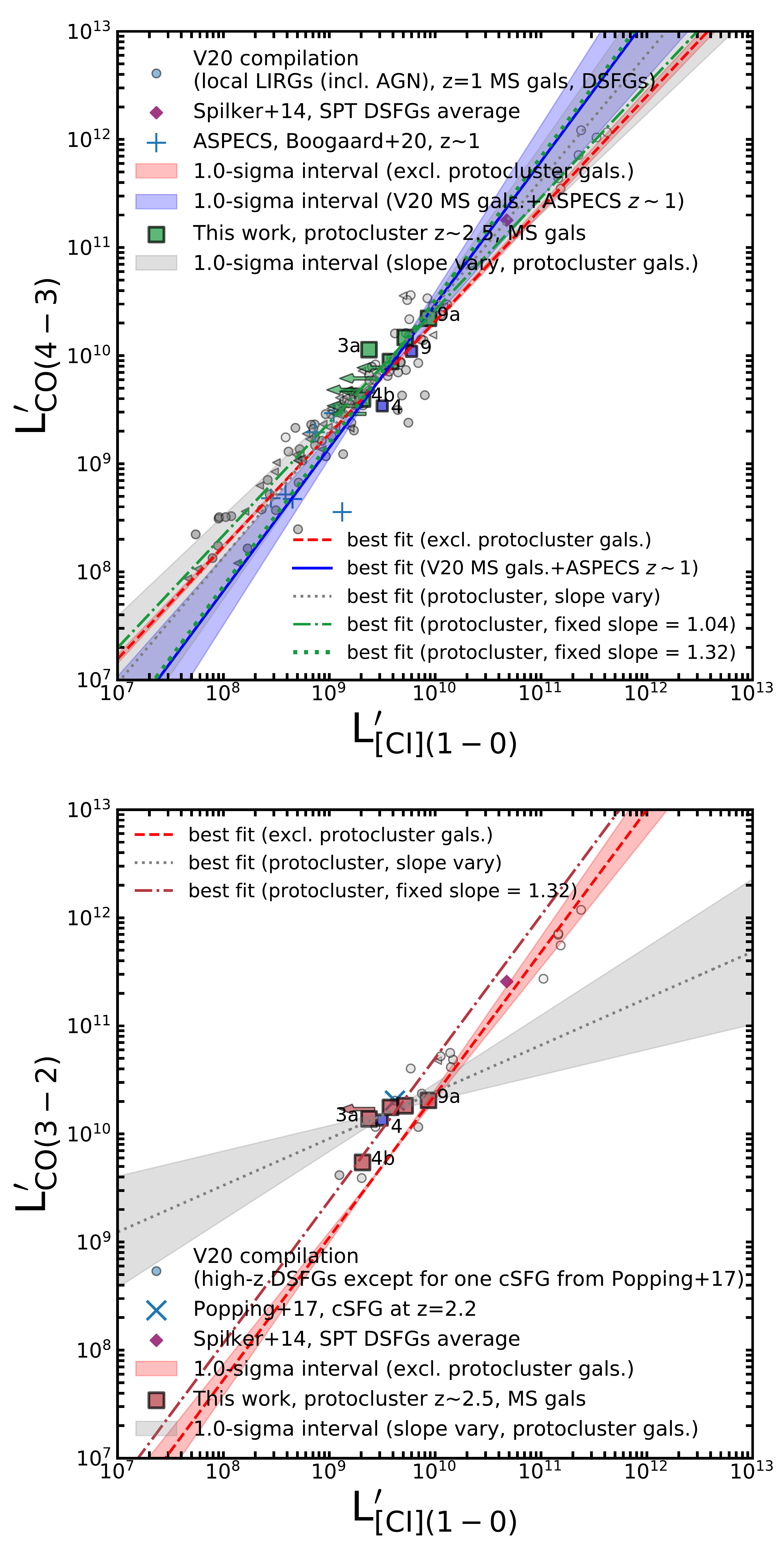}
\caption{The relations between \cione luminosity and CO (4--3) luminosity (top) and between \cione luminosity and CO~(3--2) luminosity (bottom). Protocluster galaxies are shown as squares, and smaller squares (HAE~4 and ~9) are excluded in the gas mass calibration in Section~\ref{sec:calmass} (see the text and Table~\ref{tab:lineflux} caption for the description of the galaxy name tags). We also plot other data points from \citetalias{Valentino2020b} (circles), \citet{Spilker2014} (diamond), \citet{Boogaard2020} (crosses), and \citet{Popping2017} (X-cross). In each panel, the 3$\sigma$ upper limits for [CI] non-detection with CO line detections are also plotted; arrows for the protocluster members and left faced triangles for others. Red dashed and blue solid lines are the best-fit using the data points excluding the protocluster galaxies (see Table~\ref{tab:lumifit} and the main text). Using six secure detections (HAE~3a, 4b, 8, 9a, 10 and 16) with at least two line detections, dash-dotted and dotted lines are the best-fit depending on the model assumptions. Shaded regions indicate 1$\sigma$ uncertainties of the best-fit parameters for varying slopes.\label{fig:luminosity}}
\end{figure}

\subsection{The $L'_{\rm [C I](1-0)}$ -- $L'_{\rm CO(J-J-1)}$ relation}\label{sec:lumiratio}

\begin{deluxetable*}{ccccccc}
\tablecaption{$L'_{\rm CO}$ and $L'_{\rm CI}$ relation fit\label{tab:lumifit}}
\tabletypesize{\footnotesize}
\tablewidth{0.99\textwidth}
\tablehead{
\colhead{} & \multicolumn{3}{c}{$\log{L'_{\rm CO43}} = a \times (\log{L'_{\rm CI10}}-10) + b$}& \multicolumn{2}{c}{$\log{L'_{\rm CO43}} = a_{\rm fixed} \times (\log{L'_{\rm CI10}}-10) + b$}\\
 \colhead{References} & \colhead{(1)+(2)+(3)} & \colhead{high-$z$ MS gals in (1), and (2)}& \colhead{protocluster (this work)} &\multicolumn{2}{c}{protocluster (this work)}}
\startdata
a	&	$1.04\pm0.02$		& $1.32\pm0.22$ 	&	$1.16\pm0.23$	&	fixed to $1.04$ & fixed to $1.32$\\ 
b	&	$10.32\pm0.02$	& $10.47\pm0.12$	&	$10.46\pm0.08$ & 	$10.42\pm0.03$ & $10.51\pm0.03$\\
\hline
\\
\hline\hline
{} & \multicolumn{3}{c}{$\log{L'_{\rm CO32}} = a \times (\log{L'_{\rm CI10}}-10) + b$} & \multicolumn{2}{c}{$\log{L'_{\rm CO32}} = a_{\rm fixed} \times (\log{L'_{\rm CI10}}-10) + b$}\\
References& \multicolumn{2}{c}{(1)+(3)\tablenotemark{$*$}}  & protocluster (this work) &\multicolumn{2}{c}{protocluster (this work)}\\
\hline
a&	\multicolumn{2}{c}{$1.32\pm0.10$}		&	$0.43\pm0.20$ &	\multicolumn{2}{c}{fixed to $1.32$}\\
b&	\multicolumn{2}{c}{$10.36\pm0.05$}		&	$10.39\pm0.08$&	\multicolumn{2}{c}{$10.70 \pm 0.08$}\\
\enddata
\tablecomments{
References (1) = \citetalias{Valentino2020b}, (2) = \citet{Boogaard2020}, (3) = \citet{Spilker2014}\\
$*$: We note that for this fitting, the sample is largely dominated by high-$z$ DSFGs and only one is a main-sequence galaxy at $z=2.2$ from \citet{Popping2017} that observed a compact star-forming galaxy.\\
The fit for the protocluster is using six secure galaxies at least two line detections (i.e., HAE~3a, 4b, 8, 9a, 10 and 16). If we include other \cione \,3$\sigma$ upper limits of the remaining galaxies for CO~(4--3) (see Section~\ref{sec:reanalysis} and Table~\ref{tab:lineflux} note regarding single line detections), the best-fit parameters change slightly but still consistent within the uncertainties; slope for CO~(4--3) becomes steeper with higher normalization, i.e., $a=1.40 \pm 0.26$, $b = 10.55 \pm 0.13$; if we fix the slope, $b= 10.39\pm0.04$ and $10.52 \pm 0.04$ for $a=1.04$ and $1.32$, respectively.}
\end{deluxetable*}

We compare the line luminosities between the \cione(1--0) and CO lines.
We first fit the literature values compiled in \citetalias{Valentino2020b} and $z\sim1$ galaxies from the ASPECS survey (\citealt{Boogaard2020}) as well as the mean values of the SPT sources from \citet{Spilker2014}.
The data from \citetalias{Valentino2020b} includes local IR-luminous galaxies (LIRGs), high-$z$ main-sequence galaxies, where most of the samples are at $z\sim1$ except for one at $z=2.2$, and high-$z$ dusty star-forming galaxies (DSFGs) and QSOs. We note that the local galaxies compiled in \citetalias{Valentino2020b} are representative of star-bursting population rather than typical spiral galaxies.
The mean deviation of these local galaxies from the main-sequence defined by \citet{Speagle2014} is $\langle\log\Delta {\rm MS}\rangle=1.3$. Hereafter, we define ``starburst'' galaxies to be galaxies well-above the main-sequence by 3$\sigma$ from the main-sequence (i.e., $\log{\Delta{\rm MS}}>0.6$).

We fit the data using the orthogonal distance regression to take into account uncertainties in both axes.
For fitting, we have included the non-detections using the 3-$\sigma$ upper limits if another line is detected, and the uncertainties of all data points are set to the 1-$\sigma$ noise level. 
We use a linear model of $\log{L'_{\rm CO43\,,or\,CO32}} = a \times (\log{L'_{\rm CI10}}-10) + b$.
We fit for $L'_{\rm CO43}-L'_{\rm CI10}$ relation by excluding star-bursting population of local galaxies and high-$z$ DSFGs (i.e., $z\simeq1-2$ main-sequence galaxies only) as well.
For $L'_{\rm CO32}-L'_{\rm CI10}$ relation, the data sets become sparse and most of the galaxies available are high-$z$ DSFGs except for one from \citet{Popping2017}, which is a compact star-forming galaxy (cSFG) at $z=2.2$.

The best-fit slope values indicate super-linear relations between $L'_{\rm CO}$ and $L'_{\rm [CI](1-0)}$. 
{A summary of the best-fit parameters is listed in Table~\ref{tab:lumifit}. Figure~\ref{fig:luminosity} shows the best-fit relations (red dashed -- all galaxies in the literature -- and blue solid lines -- high-$z$ main-sequence galaxies --).
We also note that the residual variance is higher for the CO~(3--2)--[C I] fit, partially owing to the limited data points. Simply by taking the best-fit models listed in the second column in Table~\ref{tab:lumifit}, i.e., (1)+(2)+(3) fit and (1)+(3) fit for  $L'_{\rm CO43}-L'_{\rm CI10}$ and  $L'_{\rm CO32}-L'_{\rm CI10}$, respectively), the brightness temperature (luminosity) ratio between CO~(4--3) and CO~(3--2) ($R_{43} = L'_{\rm CO43}/L'_{\rm CO32}$) becomes higher (smaller) for smaller (higher) $L'_{\rm [CI](1-0)}$ luminosity. 

We perform the same exercise for the protocluster galaxies only while keeping in mind that the dynamic range is much narrower i.e.,  less than an order of magnitude in luminosities.
We fit using six galaxies with secure detections (i.e., at least two lines of detection; HAE~3a, 4b, 8, 9a, 10 and 16). 
The best-fit parameters are summarized in Table~\ref{tab:lumifit} using the same fitting model but with two options of a variable slope and a fixed slope.

For $\log{L'_{\rm CO(4-3)}}-\log{L'_{\rm [CI](1-0)}}$ relation, the slope is super-linear for the protocluster members, consistent with other best-fit using the literature values.
The super-linear relation between $\log{L'_{\rm CO(4-3)}}$ and $\log{L'_{\rm [CI](1-0)}}$ may indicate that CO~(4--3) emission is not tracing the total gas mass, but either dense gas that directly forms stars and, hence star-formation, while the \cione emission is tracing the total gas mass, mimicking the global Kennicutt-Schmidt law (\citealt{Schmidt1959, Kennicutt1998a}).
Also, it may be worth noting that there is a hint of a steeper super-linear slope for high-$z$ galaxies than our protocluster galaxies.
Concerning this, there are a few studies in local galaxies (e.g., \citealt{Juneau2009, Bayet2009a}) that reported a super-linear correlation between far-infrared luminosity and low-$J$ CO transitions  ($J<3$) spanning a wide IR luminosity range.
Some local galaxy surveys reported a linear relation instead (e.g., \citealt{Gao2004, Greve2014}) for $L_{FIR}>10^{11}\,L_{\odot}$ galaxies.
The difference is often explained as the different dense gas fractions (i.e, constantly higher dense gas fraction for galaxies with high IR luminosity; e.g., \citealt{Gao2004, Wu2005, Greve2014}).
\cite{Greve2014} discussed further for the $L_{\rm FIR}$-low-$J$ CO relation that a linear slope or higher normalization factor may be physically related to the existence of dense gas, which is limited by the Eddington limit (see also \citealt{Andrews2011}).
If we can apply a similar interpretation, a hint of a shallower slope (close to a linear relation) and a higher normalization factor in protocluster galaxies than other high-$z$ main-sequence galaxies might hint at the existence of more denser gas in protocluster galaxies than high-$z$ field main-sequence galaxies.

There is a hint of a systematically different relation in the protocluster for the $L'_{\rm CO(3-2)}-L'_{\rm [CI](1-0)}$ relation (at $1\sigma$). 
Considering that the fit for the literature values is biased towards high-$z$ DSFGs and the fit for protocluster galaxies with a varying slope shows a sub-linear relation, our previous simple interpretation of $L'_{\rm CO}-L'_{\rm [CI](1-0)}$ as an analog of $L_{\rm FIR}-L'_{\rm CO}$ (or the K-S law) might not be always viable. 
Or, the CO excitation condition might be different from high-$z$ DSFGs while CO~(3--2) line still traces dense gas. 
Regarding this, it may be worth noting that the line ratio ($L'_{\rm CO(3-2)}/L'_{\rm [CI](1-0)}$) of cSFG from \citet{Popping2017} is comparable to the protocluster galaxies.
In any case, it is a surprising result to see such a large difference in these two close transitions but in a (similarly) narrow dynamic range.

The current data set is not sufficient to robustly conclude because of the the low number statistics and unknown systematic biases for currently available data sets in different environments. 
Depending on our model selection for the fit (fixed slope or not), the relation between $L'_{CO43}/L'_{CO32}$ as a function of $L'_{\rm CI10}$ also varies (negative or positive), which further denotes the current limitation. 
Therefore, we defer further discussions on the existence of systematically $L'_{\rm CO}-L'_{\rm [CI](1-0)}$ relations in different environments to a future work. We will have to wait for more data from future surveys for fairer comparison. 

\subsection{Molecular gas mass from \cione, CO and dust}
\begin{deluxetable}{cccc}
\tablecaption{Derived gas mass based on \cione, CO and dust\label{tab:gasmass}}
\tabletypesize{\footnotesize}
\tablewidth{0.99\textwidth}
\tablehead{
\colhead{ID} & \colhead{$M_{\rm mol, CI}$}  & \colhead{$M_{\rm mol, CO}$} & \colhead{$M_{\rm mol, RJ}$}\\
\colhead{} 	& {$\times 10^{10}\,M_{\odot}$}	& {$\times 10^{10}\,M_{\odot}$}	& {$\times 10^{10}\,M_{\odot}$}} 
\startdata
HAE3a       &   5.5$\pm$3.0 &   11.4$\pm$3.2    &   8.0$\pm$2.7\\
HAE4     &   7.4$\pm$2.2   &   11.2$\pm$2.7   &   $<$ 4.4\\
HAE4b     &   4.8$\pm$3.0   &   4.5$\pm$1.9   &   $<$ 4.4\\
HAE8        &   8.9$\pm$1.5 &   14.6$\pm$2.4    &   9.4$\pm$2.0\\
HAE9a       &   20.3$\pm$5.2    &   16.9$\pm$2.6    &   15.9$\pm$3.0\\
HAE10     &   $<$ 6.3   &   14.1$\pm$3.2   &   5.2$\pm$1.7\\
HAE16     &   12.2$\pm$2.6   &   15.1$\pm$3.7   &   $<$ 9.8\\
\hline
\enddata
\end{deluxetable}

\begin{figure}[t]
\centering
\includegraphics[width=0.48\textwidth, bb=0 0 1000 1000]{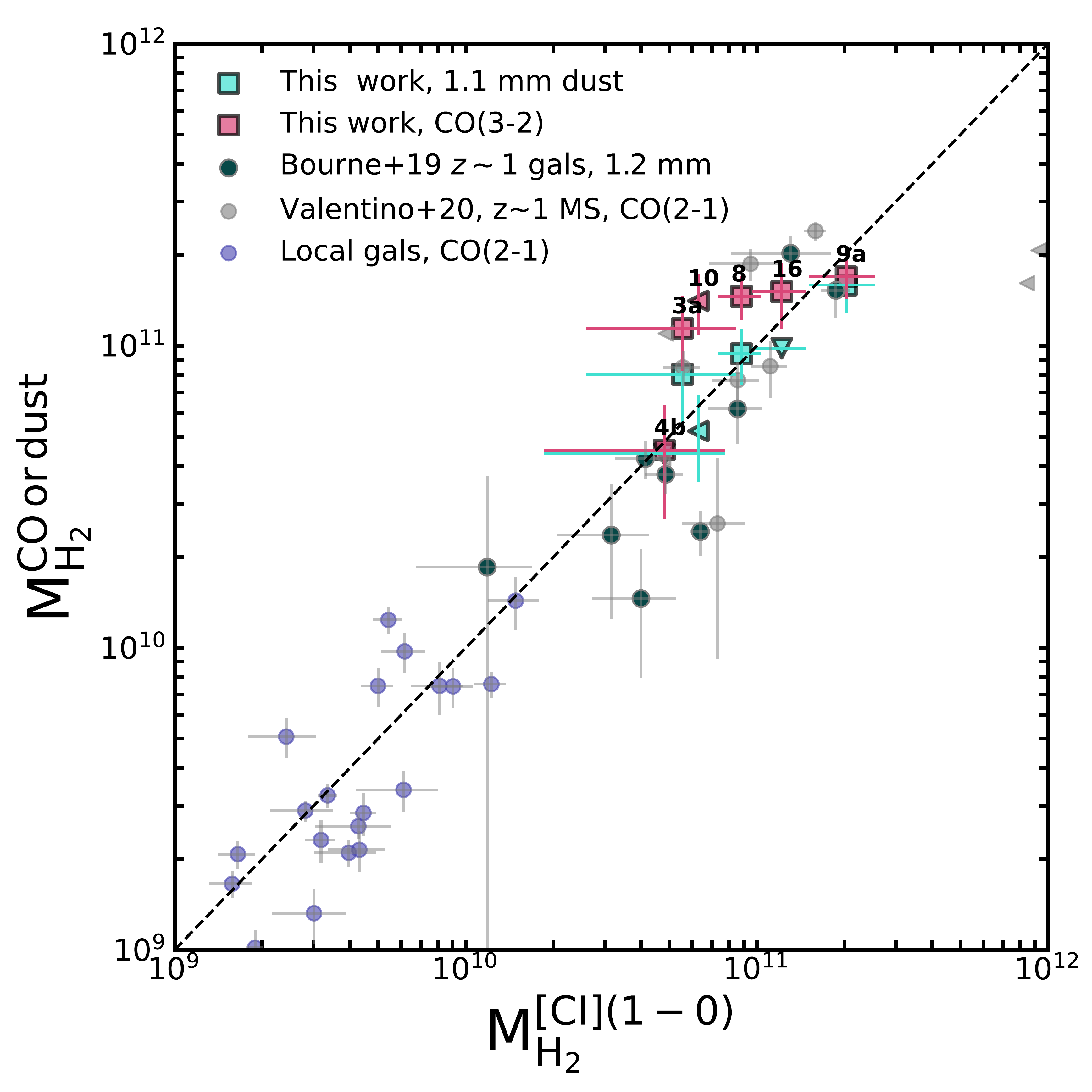}
\caption{Comparison between the derived molecular gas mass from \cione, CO and dust. Squares show the measurements from this work using 1.1 mm (brown) and CO~(3--2) (turquoise). Upper limits (3$\sigma$) are plotted as triangles. We compare our results with $z\sim1$ galaxies on the main-sequence and local galaxies compiled in \citetalias{Valentino2020b}, where both CO~(2--1) and \cione fluxes are available. For a comparison of dust-based measurements, we plot data points from \citet{Bourne2019} using the 1.2mm detections. The dashed line is to guide 1:1 relation. \label{fig:gasmass}}
\end{figure}

The relation between CO and \cione demonstrates that \cione traces the global gas mass well.
We derive gas mass from \cione, CO and dust to investigate whether the general assumptions are valid to protocluster galaxies.

Following \citet{Papadopoulos2004a}, we estimate the total molecular gas mass from \cione as
\begin{equation}
\frac{M^{\rm [CI]}_{\rm mol}}{M_{\odot}} = \frac{4 \pi \mu m_{\rm H_2}}{h\,c\,A_{10} \,X_{\rm CI} \, Q_{10}} \left( \frac{d^2_{\rm L}}{1+z}\right) S'_{\rm [CI]} \, \Delta V
\end{equation}
where $\mu = 1.36$ accounts for the mass of helium and $m_{\rm H_2}$ is the molecular mass of H$_2$ in $M_{\odot}$. 
For the \cione(1--0) transition, the Einstein coefficient A$_{10}$ is $7.93\times10^{-8}$ s$^{-1}$ and the excitation factor $Q_{10}$ is determined by the gas density and temperature. 
Without measurements of both \cione(1--0) and \cione(2--1), we cannot measure $Q_{10}$.
Here, we adopt a moderate value of $Q_{10} = 0.35$ for typical conditions of kinematic temperature $20<T_{\rm kin}<40$ K and density $300<n<10^4$ cm$^{-3}$ based on \citet{Papadopoulos2012b} and \citet{Jiao2017}.
$S'_{\rm [CI]} \, \Delta V$ is the measured \cione flux in Jy km s$^{-1}$. The remainders are luminosity distance $d_L$, speed of light $c$, and Planck constant in SI unit.
To convert the \cione mass into the H$_2$ mass, we need to assume $X_{\rm CI}$.

Observational constraints of the $\log{(X_{\rm CI})}$ for the Milky Way are in a range of [-5.7, -4.7] (\citealt{Frerking1989}) and the typically assumed value is -4.5 (\citealt{Weiss2003}).
For local galaxies without AGNs, based on the comparison with low $J$ CO-based gas measurements, they have higher abundance ratio of $\log{(X_{\rm CI})} = -4.2\pm0.2$ and comparable to high-$z$ DSFGs ([-3.9, -4.2], \citealt{Walter2011, Danielson2011, Alaghband-Zadeh2013, Valentino2018}).
Here, we adopt the mean value of $\log(X_{\rm CI}) = -4.8$ found in \citet{Valentino2018} that studied main-sequence galaxies at $z\sim1$. 

For CO~(3--2) and dust, we use the same recipe adopted in \citetalias{minju2017a}.
In brief, we take a line luminosity ratio between CO~(3--2) and CO~(1--0) ($R_{13}$ = 1.9), a similar value measured or used in \citet{Dannerbauer2009}, and \citet{Tacconi2018} for $z=1-3$ main-sequence galaxies, but see e.g., \citet{Bolatto2015, Riechers2020, Boogaard2020} for a lower value, or a slightly higher average value reported in \citet{Daddi2015}.
We use the Milky Way (MW)-like CO-to-H$_2$ conversion factor ($\alpha_{\rm CO} = 4.36$~M$_{\odot}$ (K km s$^{-1}$ pc$^{2}$)$^{-1}$; e.g., \citealt{Bolatto2013, Genzel2015}).
Here, we don't correct for the metallicity-dependence of the $\alpha_{\rm CO}$ conversion factor.
For the underlying stellar mass range that we observe ($\log{(M_{\rm star}/M_{\odot})}>10.5$), the metallicity-dependent $\alpha_{\rm CO}$ factor only increases the gas estimate the up to 0.07 dex for the lowest mass galaxies (HAE~16 and HAE~10) and does not affect higher mass regime.
For dust-based measurements, we adopt the calibration described in \citet{Scoville2016}.
The details of the calculations are presented in \citetalias{minju2017a}.

Figure~\ref{fig:gasmass} compares the molecular gas masses derived from three different tracers, and Table~\ref{tab:gasmass} summarizes the derived values.
The gas mass derived from CO is different by -0.08 dex up to 0.35 dex compared to the \cione-based gas mass and 0.14 dex higher on average.
Meanwhile, the dust-based measurements are lower than the \cione-based mass by 0.02 dex on average, and the differences range between -0.11 dex and 0.16 dex.
The mean values of the difference compared to the \cione-based measurement are shown as the arrows along the vertical line of the top-left panel of Figure~\ref{fig:dev}.
In summary, three gas tracers are broadly in good agreement for the inferred molecular masses within $\approx$0.3 dex, but CO-based gas measurements give $\approx$40\% (0.14 dex) higher value than other gas tracers on average.

For comparison, we also plot the local and high-$z$ field \cione studies in Figure~\ref{fig:gasmass}.
The data points are taken from \citetalias{Valentino2020b} and \citet{Bourne2019}.
\citetalias{Valentino2020b} collected local galaxy and high-$z$ \cione(1--0) studies.
From their catalogs, we take the $z\sim1$ main-sequence measurements and local galaxies for which CO~(2--1) measurements are available.
For the $z\sim1$ galaxies, we assumed $R_{\rm 12} = 1.2$ and the same $\alpha_{\rm CO}$ and $X_{\rm CI}$ with protocluster galaxies given that they are on the main-sequence.
For local galaxies with CO~(2--1) we used the different $\alpha_{\rm CO} = 0.8$ and $\log{(X_{\rm CI})} = -4.2$ given that they are on average above the main-sequence ($\langle\log\Delta {\rm MS}\rangle$=1.3).
From \citet{Bourne2019}, we compare the dust-based measurements and \cione.
From the Table~2 in \citet{Bourne2019}, we used the flux values obtained from larger aperture sizes for Galaxy 1 and 2.
In general, they are well-matched with each other.

\section{Discussion} \label{sec:discussion}
\subsection{Self-consistent calibration from three tracers}\label{sec:calmass}

\begin{figure*}[t]
\centering
\includegraphics[width=0.85\textwidth, bb=0 0 1000 1000]{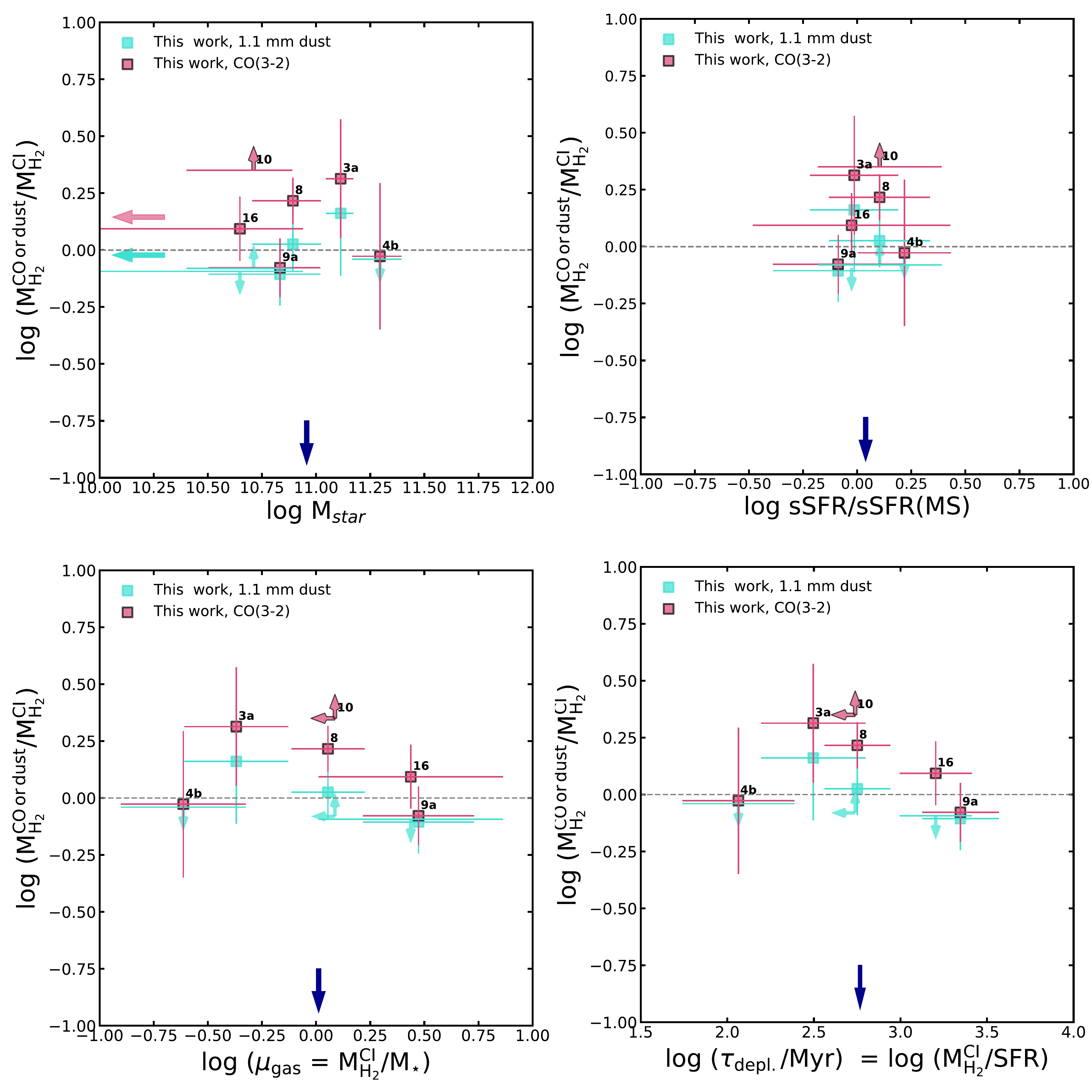}
\caption{The gas mass comparison from CO~(3--2), dust and \cione(1--0) emissions. In the vertical axes, we plot the CO- and dust- based gas mass divided by \cione-based mass. In the horizontal axes, we plot stellar mass (top left), deviation from the main-sequence (top right), molecular gas-to-stellar mass ratio (bottom left) and depletion time scale (bottom right). The mean values of the galaxy properties are marked as dark blue arrows. The arrows plotted along the vertical axis of the top left panel is the average deviation of dust- based (turquiose, -0.02 dex) and CO-based (pink, +0.14 dex) from \cione- based including the upper limits.\label{fig:dev}}
\end{figure*}

The general key assumptions for deriving gas mass from different tracers are the following.
In the dust-based calibration of \citet{Scoville2016}, the mass-weighted dust temperature is fixed to $T_{\rm d}$ = 25 K and the dust-to-gas mass ratio is implicitly fixed.
In the CO-based calibration, we need to assume the $\alpha_{\rm CO}$ and CO gas excitation condition.
For the \cione-based calibration, we need to assume the $X_{\rm CI}$ abundance ratio (or \cione-conversion factor).
The excitation temperature may matter, especially for the higher transition \cione~(2--1) but has less impact for \cione~(1--0) (\citealt{Weiss2005}). 
The value of $X_{\rm CI}$ is not constrained observationally in high-z galaxies.
In the following, we discuss the possible changes of the parameters ($T_{\rm d}$, gas-to-dust mass ratio, $\alpha_{\rm CO}\times R_{13}$ and $X_{\rm CI}$) to get self-consistent gas mass from different tracers focusing on non-detection case.

If the dust continuum is non-detection (i.e., HAE~4 and HAE~16), it may be attributed to a high dust temperature and/or intrinsically low dust mass (at fixed $T_{\rm d}$). For a case of high dust temperature, the dust SED would shift to a shorter wavelength hampering the detection at a given sensitivity.
It would imply the dust-based gas mass to an even lower value if the mass-weighted temperature is also high enforcing other gas measurements also to be lowered.
Alternatively, if the dust mass intrinsically is low, this would require higher gas-to-dust mass ratio to match other gas measurements.
The dust-to-gas mass ratio is closely connected to the metallicity (e.g., \citealt{Remy-Ruyer2014}).
Considering the mass-metallicity relation and the stellar mass of the galaxies, e.g., $\log{(M_{\rm star}/M_{\odot})} \approx 11.3$ (HAE4) and $\approx 10.6$ (HAE16), the scatter of the gas-to-dust ratio at fixed metallicity ($\sim0.37$ dex) could allow such variation.
Therefore, we can reasonably choose the 3$\sigma$ upper limit as face value for the following discussion.

Firstly, we consider whether the mild differences are depending on galaxy parameters, especially for the CO-based measurements, which show systematically higher value than the \cione-based gas mass.
Figure~\ref{fig:dev} compares the difference between the molecular gas masss derived from three different tracers as a function of stellar mass ($M_{\rm star}$), deviation from the main-sequence (sSFR/sSFR(MS)), gas-to-stellar mass ratio ($\mu_{\rm gas} = M_{\rm gas}/M_{\rm star}$) and gas depletion time scale ($\tau_{\rm depl.} = M_{\rm gas}/{\rm SFR}$) where the latter two are measured using the \cione-based gas measurements.
We do not find a statistically meaningful correlation with galactic parameters and the gas mass discrepancies. 
For example, we cannot reject the null hypothesis that depletion time scale and the gas mass ratio of CO-based and \cione-based measurements, which is a proxy for the overestimated amount of $\alpha_{\rm CO}\times R_{13}$ at fixed $X_{\rm CI}$, are not correlated (i.e., Spearman's Rank coefficient of = -0.37 with a $p$-value of 0.47, but note that if HAE~4b is excluded, the strength of the correlation becomes stronger, -0.90 with a moderate $p$-value of 0.04.)
Given the small number statistics, we conclude that there is no clear dependency on the stellar mass (for $\log{M_{\rm star}/M_{\odot}} = [10.5,11.3]$), deviation from the main-sequence (for $\log{\Delta \rm MS} = [-0.1,0.2]$), gas-to-stellar mass ratio (for $\log{\mu_{\rm gas}} = [-0.6,0.5]$) and depletion time scale (for $\log({\tau_{\rm depl.}}$/Myr) $= [2.1, 3.3]$), and the differences are coming from galaxy-to-galaxy variations creating a scatter.

Detailed investigations for the gas mass cross-calibration for individual galaxies are presented in Appendix~\ref{app:crosscalibration}. We briefly summarize here the results as follows:
(1) the CO~(4--3)-to-[C I](1--0) luminosity ratio of two galaxies (HAE~3a and HAE~16) suggests the existence of dense gas that may resemble the ISM properties of DSFGs (see also Section~\ref{sec:lumiratio});
(2) the non-detection of [C I](1--0) in HAE~10, the current constraints on mass-weighted dust temperature and $\log{(X_{\rm CI})}$ suggest that it is more likely that the assumed $\alpha_{\rm CO}\times R_{13}$ would be lower than what is assumed. 
We note all of these galaxies are all on the main-sequence defined by \citet{Speagle2014} (i.e., $\log{\Delta}$MS $\leq \pm0.2$).

Overall, there is a hint that the assumed $\alpha_{\rm CO}\times R_{13}$ needs to be lowered for HAE~3, 10, and 16 (thus reducing the CO-based gas mass), though it is not conclusive for the remaining galaxies. If the protocluster galaxies share similar properties, it is more reasonable to lower the CO-based gas mass (thus reducing the $\alpha_{\rm CO}\times R_{13}$) to match with dust and \cione based gas mass.
We finally note that the gas mass cross calibrations do not require more than a factor of $>0.6$ dex change that would characterize the protocluster galaxies well above the main-sequence where the assumptions of $\alpha_{\rm CO}\times R_{\rm 13}$ and $X_{\rm CI}$ are different.

\subsection{Gas depletion time scale compared to field galaxies}

\begin{figure*}[t]
\centering
\includegraphics[width=0.98\textwidth, bb=0 0 1500 500]{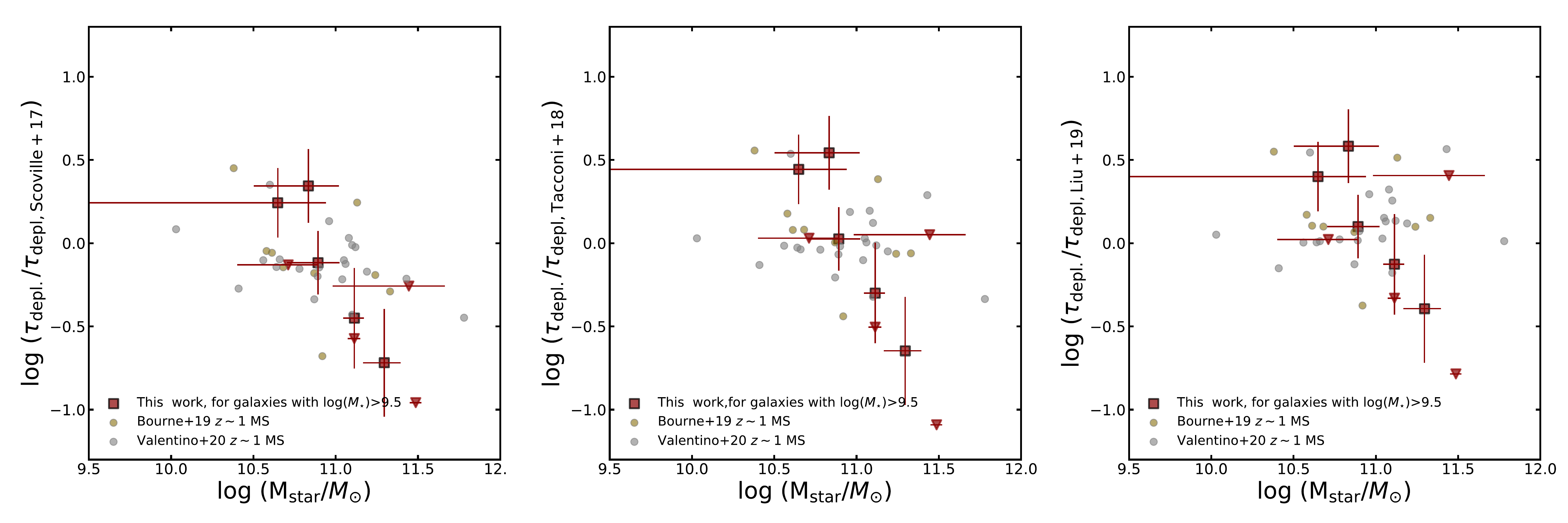}
\caption{Gas depletion time scale for the protocluster galaxies as a function of stellar mass normalized by field scaling relation (depletion time scale) from \citet{Scoville2017b} (left), \citet{Tacconi2018} (middle) and \citet{Liu2019b} (right). In addition to five \cione detections (squares), non-detections for galaxies above $\log{(M_{\rm star}/M_{\odot})}>9.5$ are also shown as up-side-down triangles. Filled circles are other \cione-based data points for $z\sim1$ main-sequence for a reference of field galaxies from \citet{Bourne2019} and \citetalias{Valentino2020b}. \label{fig:depl}}
\end{figure*}

In the previous section, we showed that the \cione line is a reliable gas tracer of protocluster galaxies.
The same calibration of $\log(X_{\rm CI}) = -4.8$ applied for the field main-sequence galaxies can be used for these galaxies.
By investigating \cione-based gas mass, we gained one additional data point compared to 1.1 mm dust and more robust gas measurement compared to CO.
In this section, we revisit how gas scaling relation (i.e., gas depletion time scale determined by redshift, stellar mass, and deviation from the main sequence) differs from field galaxies based on the \cione observations.

Figure~\ref{fig:depl} shows the gas depletion time scale of the protocluster galaxies normalized by the (field) gas scaling relation determined in \citet{Scoville2017b}, \citet{Tacconi2018} and \citet{Liu2019b} at given stellar mass, redshift and deviation from the main sequence.
We plot for the \cione non-detections as well for those with CO~(4--3) line detections for galaxies with $\log{M_{\rm star}/M_{\odot}}>9.5$, where the stellar mass constraint is more reliable than the lower mass regime.
For non-detections, we assume the \cione line width the same as the CO~(4--3) width (the sixth column of Table~\ref{tab:lineflux}) to get the 3$\sigma$ upper limit of the depletion time scale.
Other \cione-based data points in circles are $z\sim1$ main sequence galaxies taken from \citet{Bourne2019} and \citetalias{Valentino2020b}.

Compared to high-$z$ ($z\sim1$) field main-sequence galaxies, we find a trend that the depletion time scale for main-sequence protocluster galaxies steeply decreases with stellar mass.
In comparison with field scaling relation, a stronger anti-correlation is found between depletion time scale normalized by three different scaling relations and stellar mass in protocluster galaxies.
We find negative Spearman coefficient values ($p$-value) of -0.81 (0.008, \citealt{Scoville2017b}), -0.68 (0.04, \citealt{Tacconi2018}), -0.53 (0.14, \citealt{Liu2019b}), when the 3$\sigma$ upper limits are all taken into account.
We also note that the correlation coefficient is -0.9 with p-value of 0.04 if only the secure detections are considered.
As control samples, field galaxies from \citet{Bourne2019}, and \citetalias{Valentino2020b} show less compelling evidence of such anti-correlation, i.e., Spearman coefficient ($p$-value) = -0.35 (0.06, \citealt{Scoville2017b}), -0.20 (0.30, \citealt{Tacconi2018}) and 0.23 (0.22, \citealt{Liu2019b}).
The gas scaling relation in \citet{Liu2019b} has the strongest stellar-mass dependency compared to the other two studies, which is why we find less significant signature of anti-correlation when normalized by \citet{Liu2019b} in the protocluster members and field galaxies, if any.

With the \cione line detections, we reconfirm the trend seen in \citetalias{minju2017a} and \citet{Tadaki2019a} that used CO/dust and CO only, respectively, though the absolute vertical scale would be slightly different due to the different flux values.
Overall, we find a signature of stronger anti-correlation between the stellar mass and the gas depletion time scale normalized by field scaling relation: the more the galaxy becomes massive, the faster it consumes or removes gas, though we caution that the number statistics and potential selection effects are remaining concerns.

What kind of physical mechanisms can derive this trend? 
In \citetalias{minju2019a}, we discussed that the larger CO~(4--3) line widths for protocluster galaxies could arise from unresolved gas-rich mergers or smaller sizes.
These are not necessarily exclusive that mergers can also make galaxies smaller.
There are still other possible mechanisms to make the galaxy smaller, for example, compaction through disk-instability (e.g., \citealt{Bournaud2008, Genzel2008, Genel2012, Agertz2009, Ceverino2010, Zolotov2015, Tacchella2016, Barro2016}) and gas stripping (e.g., \citealt{Gunn1972,  Haynes1984, Poggianti2017}).
Gas stripping in molecular phase remains inconclusive (e.g., \citealt{Stark1986, Kenney1989, Boselli1997}), but there is a growing body of evidence that more tightly bound molecular gas can even be stripped or disturbed (e.g., \citealt{Vollmer2008, Fumagalli2009, Boselli2014a, Zabel2019}).
We note also that the conventional picture of gas stripping is more likely to happen in low mass galaxies in massive clusters, but some group members also show such signature for neutral atomic gas (e.g., \citealt{VerdesMontenegro2001, Rasmussen2006, Hess2013}).
Considering these, the molecular gas stripping in massive galaxies in the group-like environments (i.e., protoclusters) may be less efficient, but it remains one of the possibilities.
At the same time, this protocluster also shows a hint that more massive galaxies tend to reside in denser regions (\citetalias{minju2017a}) that other environmentally driven mechanisms (gas strangulation; \citealt{Larson1980,  Peng2015}) might play an additional role for the protocluster members, for earlier quenching.
All these mechanisms can be efficient in consuming gas or removing gas out of the galaxies, though the efficiency of these mechanisms in protoclusters is unknown observationally and
current data sets are not sufficient to reject or accept any of the possibilities.
In this regard, we may get more hints of possible mechanisms from either very massive ($M_{\rm star}/M_{\odot}>11$) or lower mass ($M_{\rm star}/M_{\odot}<10.5$) galaxies in protoclusters, which will also help to verify the observed trend.

\section{Summary and conclusion} \label{sec:summary}
We revisited the gas content of the protocluster by \cione line observations using ALMA. 
We aimed to test if the same calibrations in measuring gas content using CO and dust can be applied to protocluster galaxies using an independent gas tracer of \cione.
Five galaxies are detected in \cione out of sixteen galaxies targeted.
We first compared the CO-to-[C I] luminosity relation with other studies. 
It revealed the \cione is a good gas tracer for the protocluster galaxies. 
The CO~(4--3)-to-\cione line relation further hinted at the presence of denser gas in protocluster galaxies compared to field galaxies though the dynamic range was still too narrow to robustly conclude on the existence of different relation in different environments.
We then compared the gas mass based on \cione, CO(3--2), and dust when at least one of the last two tracers was available.
Given that these five galaxies are all on the main-sequence, we adopted the general calibration methods applied to (field) main-sequence galaxies by assuming $\log{(X_{\rm CI})} = -4.8$ for the \cione-based gas measurement. 
The \cione-based gas measurements are, in general, bridging the gap between CO~(3--2)-based and dust-based gas estimates.
They are lower or comparable to the CO-based gas measurements by -0.35 dex at the lowest.
The difference between \cione- and the dust- based measurements are mild by up to 0.16 with the mean difference of 0.02 dex.
Considering the CO~(4--3)-to-\cione line ratios and the parameter space allowed for typical main-sequence galaxies, 
there are three galaxies that CO-based gas mass may be overestimated than the other tracers, suggesting different calibrations (or conversion factors).
Hence, the combination of $\alpha_{\rm CO}\times R_{13}$ needs to be lowered for these protocluster galaxies by up to $\approx$ 0.35 (0.43) dex to match the \cione-(dust-)based gas measurements.
Such adjustment is still within the uncertainty of the individual calibration methods, however.

We generally reconfirm our previous findings that the mean gas fraction is comparable to field galaxies for a stellar mass range of $\log{M_{\rm star}/M_{\odot}} = [10.6, 11.3]$ ($f_{\rm gas} = M_{\rm gas}/(M_{\rm gas}+M_{\rm star}) \approx 50\%$)).
At least for these five \cione-detected galaxies, the depletion time scale decreases more rapidly with stellar mass than field galaxies that might lead galaxies to quench earlier than field galaxies.
More detailed studies on more massive galaxies ($\log{M_{\rm star}/M_{\odot}}>11$) and less massive ones ($\log{M_{\rm star}/M_{\odot}}>10.5$) would help to disentangle potential physical mechanisms that are dominant than field galaxies making the observed trend.

Our findings support the idea that optically-thin gas tracer of \cione is indeed a good tracer for probing gas content even for protocluster galaxies.
Based on the similarity between dust-based and \cione-based gas measurement, dust observations would be a more efficient tool for a larger number of galaxies to improve statistics, especially for the galaxies on the main-sequence.
However, \cione can still be an efficient tracer for high-$z$ where getting CO~(1--0) is more expensive, and for such, the line can be used as an alternative anchor for dust-based calibration.

Finally, we caution against the generalization of our findings. 
With limited number of galaxies detected in \cione (also CO and dust), it is difficult to verify whether there is any selection bias delivering such mass dependency.
Such bias could originate not only from the flux-limited galaxy selection (H$\alpha$) but also from the halo property (mass, evolutionary stage) of the protocluster.
More galaxies in the probed mass range ($\log{M_{\rm star}/M_{\odot}} = [10.5, 11]$) in differently selected protoclusters would also help to broaden our understanding of galaxy evolution in different environments.

\acknowledgments
We are immensely grateful to the anonymous referee for constructive comments. We thank Dr. Daizhong Liu for fruitful discussions and Dr. Frencesco Valentino for providing the observed data set before it became public. This paper makes use of the following ALMA data: ADS/JAO.ALMA \#2015.1.00152.S. ALMA is a partnership of ESO (representing its member states), NSF (USA) and NINS (Japan), together with NRC (Canada) and NSC and ASIAA (Taiwan) and KASI (Republic of Korea), in cooperation with the Republic of Chile. 
Y.T is supported by NAOJ ALMA Scientific Research Grant No. 2018-09B.
%

\facilities{ALMA}


\software{astropy \citep{astropy},  Scikit-learn \citep{scikit-learn}, CASA \citep{McMullin2007}
          }



\appendix
\section{Spectra for single line (CO~(4--3)) detections}\label{app:single}
We show in Figure~\ref{fig:remainder} the spectra of CO~(4--3) detection without other line dections. The spectra are obtained with aperture sizes listed in Table~\ref{tab:lineflux} centered at the CO~(4--3) peak. The shaded area is to indicate the integrating range for CO~(4--3) (light red), and the corresponding velocity range for \cione (light green) and CO~(3--2) (light brown).

\begin{figure*}[tb]
\centering
\includegraphics[width=0.80\textwidth, bb=0 0 550 850]{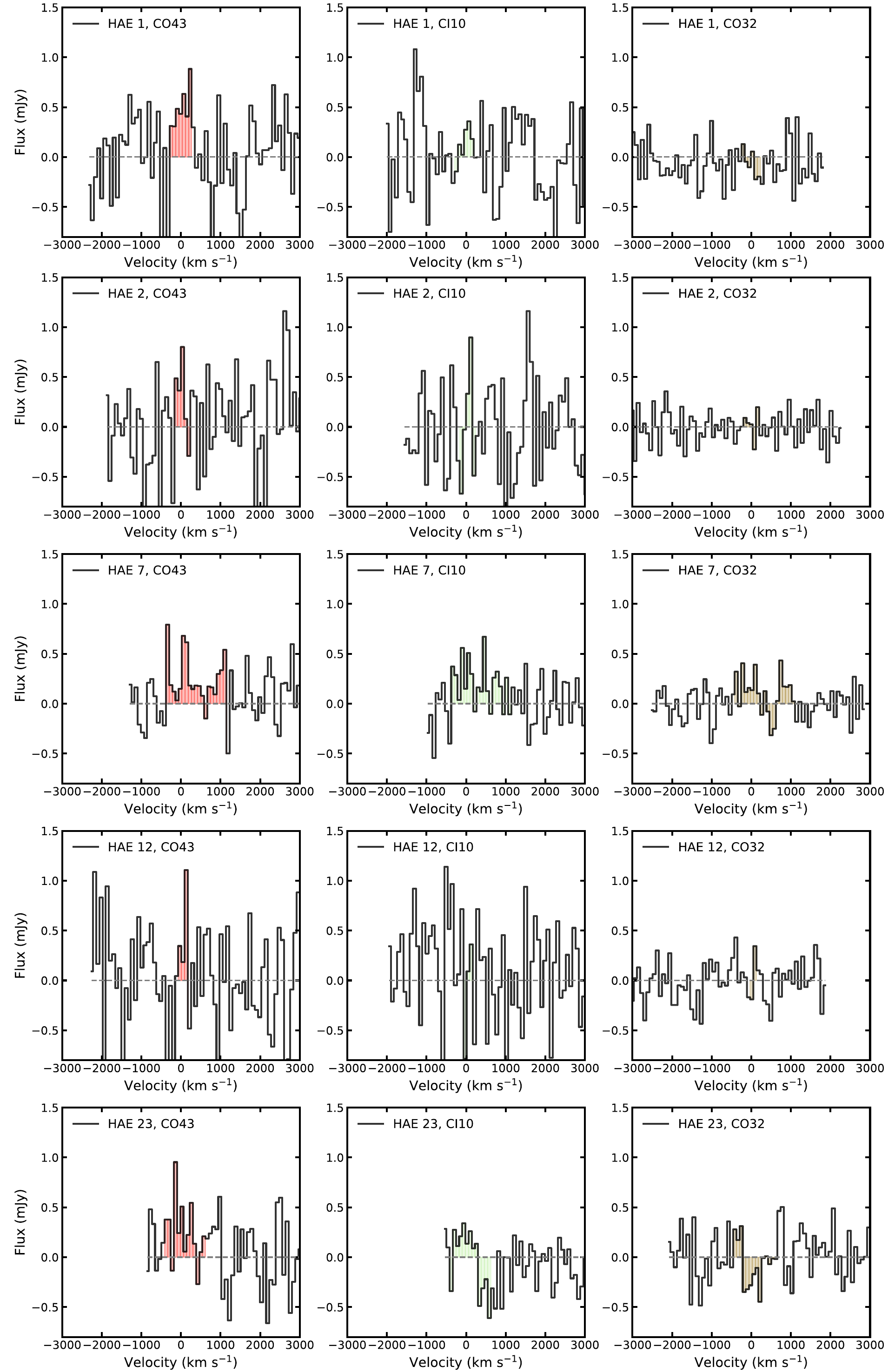}
\caption{The spectra of galaxies with CO~(4--3) single detection (see Appendix~\ref{app:single} for details).\label{fig:remainder}}
\end{figure*}

\section{HAE~4 and HAE~10 spectra at the CO~(4--3) peak position}\label{app:haes}
We show in Figure~\ref{fig:haes} the spectra of HAE~4b and HAE~10 centered at the CO~(4--3) peak position.
The shaded area is to indicate the integrating range for CO~(4--3) (light red), and the corresponding velocity range for \cione (light green) and CO~(3--2) (light brown).

\begin{figure*}[b]
\centering
\includegraphics[width=0.80\textwidth, bb=0 0 750 250]{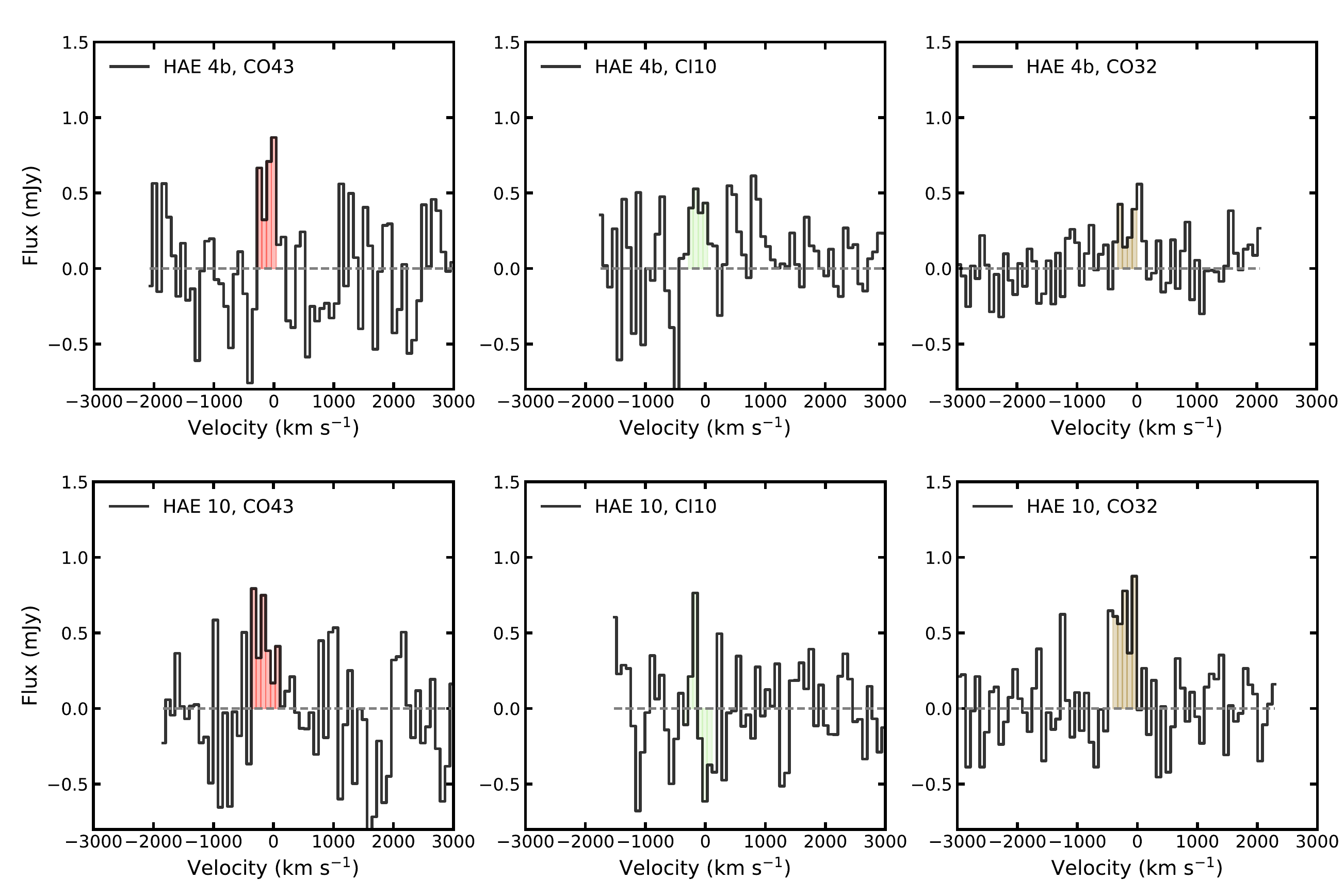}
\caption{The spectra of the HAE~4b and HAE~10 (see Appendix~\ref{app:haes} for details).\label{fig:haes}}
\end{figure*}

\section{Gas mass cross calibration of individual galaxies}\label{app:crosscalibration}

We describe here the detailed gas mass cross calibration for individual galaxies.
We investigate by first assuming that the dust measurement, which gives the lowest value on average, is true.
In this case, the $X_{\rm CI}$ is recalibrated within a range between -0.16 dex and 0.11 dex from the original assumption of $\log(X_{\rm CI}) = -4.8$, which is negligible if we consider the scatter of the reported values (e.g., \citealt{Valentino2018}, $\sigma =0.19-0.23$ dex).
For $\alpha_{\rm CO}\times R_{13}$, the adjustment could be as large as a factor of 2.7 (0.43 dex, HAE~10).

To investigate further for a best solution, we check the line luminosity ratio between [CI] and CO.
HAE~16 has a [C I]-to-CO~(4--3) line luminosity (in $L_{\odot}$) ratio ($\log{(L_{\rm CI10}/L_{\rm CO43})}$) of  $-0.37\pm0.11$.
HAE~3a shows a [C I]-to-CO~(4--3) luminosity ratio of $-0.57\pm0.17$ in log. 
These line ratios are lower relative to the average value of $z\sim1$ main-sequence galaxies, and the value is comparable to the high-$z$ DSFGs average value (\citetalias{Valentino2020b}). 
The fit to the CO-to-[CI] line luminosity (see Section~\ref{sec:lumiratio}) shows a hint of higher CO luminosity at given \cione luminosity towards higher \cione for field galaxies.
While these galaxies are not as luminous as high-$z$ DSFGs, the CO(4--3)-to-[CI] line ratios of two galaxies are comparable to them.
Using a PDR model (\citealt{Pound2008, Kaufman2006})\footnote{\url{http://dustem.astro.umd.edu/}}, the line ratio suggests the existence of $n_{\rm H}\sim10^{4.5}$ cm$^{-3}$ gas both in HAE~16 and HAE~3a.
If the ISM properties of HAE~3a and HAE~16 resembles those of DSFGs (those with lower $\alpha_{\rm CO}$ and $R_{13}$), the $\alpha_{\rm CO}\times R_{13}$ value may be lower than what is assumed, which leads to more consistent values from different tracers.
It would also be worth noting that both galaxies are on the main-sequence ($\log{\Delta \rm MS}$ =  -0.02 and -0.03 for HAE~3a and HAE~16, respectively; Figure~\ref{fig:dev} top right). 
The level of adjustment in both galaxies are -0.15 and -0.19 dex in $\alpha_{\rm CO}\times R_{13}$.

Secondly, we consider the \cione non-detection case with other emissions detected. 
In HAE~10 (see Appendix~\ref{app:haes} for CO and \cione spectra), \cione line is not detected and the dust-based gas mass is consistent with the \cione-based upper limit.
To be consistent, $\alpha_{\rm CO}\times R_{13}$ should be lowered by a factor of $\approx$2.7, otherwise, both the $X_{\rm CI}$ and dust temperature need to be recalibrated. 
The latter case is less likely, because $X_{\rm CI}$ needs to be lowered by 0.35 dex, and the dust temperature should be lower than 20 K unless the difference is due to a higher gas-to-dust ratio.
While the recipe from \citet{Scoville2016} assumes a fixed dust temperature of $T_{\rm d} = 25$ K, the mass-weighted dust temperature of star-forming galaxies could increase at least up to $z\sim4$ (\citealt{Schreiber2018}).
From the redshift-dependent dust temperature fit in \citet{Schreiber2018}, the expected dust temperature is above 30 K at $z\sim2$.
It is less likely that the dust temperature is lower than 20 K if the dust temperature evolution is only a function of redshift and happening in the same manner in the protocluster galaxies.
Meanwhile, the abundance ratio which is lower by 0.35 dex than the assumed value ($\log{(X_{\rm CI}}) = -4.8$) is not a commonly observed value for high-redshift galaxies.
The galaxy would be a unique galaxy on the main-sequence to match with CO-based measurements.
Therefore, it would be more reasonable to conclude that the CO gas excitation condition of HAE~10 is close to, but not an extreme case of, starburst-like phase lowering the $\alpha_{\rm CO}\times R_{13}$, while the galaxy is on the main-sequence by \citet{Speagle2014} ($\log{(\Delta MS)} = 0.10\pm0.29$; Figure~\ref{fig:dev} top right).


\bibliography{minjujournal}
\bibliographystyle{aasjournal}



\end{document}